\documentclass[iop, twocolappendix]{mnras} 

\usepackage[pdftex]{graphicx}

\usepackage{amsmath, amsthm, amssymb}
\let\bibhang\relax
\usepackage{natbib}
\usepackage{subfigure}
\usepackage{upgreek}
\usepackage{accents}
\usepackage{color}
\usepackage{hyperref}
\usepackage{dblfloatfix}
\usepackage[T1]{fontenc}
\usepackage{aecompl}

\newcommand{\ares}{\textsc{ares}}

\newcommand{\jwst}{{\it JWST}}
\newcommand{\edges}{{\it EDGES}}

\newcommand{\Omnow}{\Omega_{\text{m},0}}
\newcommand{\Obnow}{\Omega_{\text{b},0}}

\newcommand{\Tcmb}{T_{\gamma}}

\newcommand{\TS}{T_{\text{S}}}
\newcommand{\TK}{T_{\text{K}}}

\newcommand{\HI}{\text{H} {\textsc{i}}}

\newcommand{\Lya}{\text{Ly-}\alpha}

\newcommand{\MUV}{M_{\text{UV}}}

\newcommand{\zoff}{z_{\text{off}}}

\newcommand{\Ja}{J_{\alpha}}

\newcommand{\SFR}{\dot{M}_{\ast}}

\newcommand{\MAR}{\dot{M}_h}

\newcommand{\xibar}{\overline{x}_i}

\newcommand{\fstar}{f_{\ast}}

\newcommand{\fesc}{f_{\text{esc}}}

\newcommand{\Msun}{M_{\odot}}

\newcommand{\Tmin}{T_{\text{min}}}

\newcommand{\dTb}{\delta T_b}

\newcommand{\cXunits}{\text{erg} \ \text{s}^{-1} \ (\Msun / \text{yr})^{-1}}

\newcommand{\QHII}{Q_{\textsc{HII}}}

\title[EDGES: Implications for Galaxy Formation]{What does the first highly-redshifted 21-cm detection tell us about early galaxies?}
\author[Mirocha \& Furlanetto]{
Jordan Mirocha\textsuperscript{\thanks{mirocha@astro.ucla.edu}} and Steven R. Furlanetto
\\
Department of Physics and Astronomy, University of California, Los Angeles, CA 90024, USA\\
}

\begin{document}

\pagerange{\pageref{firstpage}--\pageref{lastpage}} \pubyear{2018}
\maketitle

\begin{abstract}
The Experiment to Detect the Global Epoch of Reionization Signature (EDGES) recently reported a strong 21-cm absorption signal relative to the cosmic microwave background at $z \sim 18$. While its anomalous amplitude may indicate new physics, in this work we focus on the timing of the signal, as it alone provides an important constraint on galaxy formation models. Whereas rest-frame ultraviolet luminosity functions (UVLFs) over a broad range of redshifts are well fit by simple models in which galaxy star formation histories track the assembly of dark matter halos, we find that these same models, with reasonable assumptions about X-ray production in star-forming galaxies, cannot generate a narrow absorption trough at $z \sim 18$. If verified, the EDGES signal therefore requires the fundamental inputs of galaxy formation models to evolve rapidly at $z \gtrsim 10$. Unless extremely faint sources residing in halos below the atomic cooling threshold are responsible for the EDGES signal, star formation in $\sim 10^8$-$10^{10} \ \Msun$ halos must be more efficient than expected, implying that the faint-end of the UVLF at $\MUV \lesssim -12$ must \textit{steepen} at the highest redshifts. This steepening provides a concrete test for future galaxy surveys with the \textit{James Webb Space Telescope} and ongoing efforts in lensed fields, and is required regardless of whether the amplitude of the EDGES signal is due to new cooling channels or a strong radio background in the early Universe. However, the radio background solution requires that galaxies at $z > 15$ emit 1-2 GHz photons with an efficiency $\sim 10^3$ times greater than local star-forming galaxies, posing a challenge for models of low-frequency photon production in the early Universe.
\end{abstract}

\section{Introduction} \label{sec:intro}
A simple picture of high-$z$ galaxy evolution has begun to emerge in recent years as the rest-frame ultraviolet luminosity function (UVLF) of galaxies has been assembled by deep surveys at $4 \lesssim z \lesssim 10$ \citep[e.g.,][]{Bouwens2015,Finkelstein2015}. The rapid decline in the abundance of bright galaxies \citep[e.g.,][]{Oesch2017}, coupled with a reduction in the measured value of the Thomson scattering optical depth of the cosmic microwave background \citep{Planck2015}, has reduced the need for an abundance of galaxies far beyond detection thresholds at $z \gtrsim 6$. If one extrapolates UVLFs to absolute magnitudes $\MUV \gtrsim -12$ (roughly the atomic cooling threshold in most models), reionization ends at $z \sim 6$, so long as the escape fraction of UV photons is $0.1 \lesssim \fesc \lesssim 0.2$ \citep[e.g.,][]{Robertson2015,Bouwens2015b}. 

21-cm observations have long been expected to provide a complementary view of structure formation in the early Universe, and thus an independent check on this emerging picture \citep{Madau1997}. Fluctuations in the background, being targeted by interferometers like PAPER \citep{Parsons2014}, MWA \citep{Dillon2014}, LOFAR \citep{Patil2017}, and HERA \citep{DeBoer2017}, trace spatial variations in the density, ionization \citep{Furlanetto2004}, kinetic temperature \citep{Pritchard2007}, and $\Lya$ \citep{Barkana2005} field, and can thus constrain the spatial distribution of sources and their radiative properties. Alternatively, the sky-averaged (``global'') 21-cm signal \citep{Shaver1999} traces the volume-averaged ionization and thermal histories in time, and thus offers a powerful independent constraint on the timing of reionization and re-heating.

Though early work predicted a global signal with only a weak high-frequency emission feature, subsequent papers found a much richer structure \citep[e.g.,][]{Furlanetto2006,Pritchard2010a,Mesinger2013} due to the inefficiency of $\Lya$ heating \citep{Chen2004,Chuzhoy2006,FurlanettoPritchard2006,Hirata2006}, which results in an IGM that remains cooler than the CMB prior to (and in some cases during) reionization. Indeed, some theoretical models in recent years predict a relatively cold IGM during reionization, if the dominant X-ray sources in high-$z$ star-forming galaxies are high-mass X-ray binaries \citep[HMXBs, as is the case locally;][]{Gilfanov2004,Mineo2012a}, since HMXBs have hard spectra \citep{Fialkov2014,Mirocha2014}, and because of the apparent inefficiency of star formation in faint galaxies, which delays the onset of heating \citep{Mesinger2016,Mirocha2017,Madau2017}. Population III stars and proto-quasars likely have a more subtle impact on the signal \citep{Mirocha2018,Tanaka2016}. For a review of high-$z$ galaxy formation and 21-cm cosmology, see, e.g., \citet{Barkana2001,FurlanettoOhBriggs2006,Pritchard2012}.

The first reported detection from the Experiment to Detect the Global Epoch of Reionization Signature \citep[EDGES;][]{Bowman2018} is consistent with a cold IGM before reionization\footnote{This result is consistent with previous limits from PAPER \citep{Parsons2014,Pober2015}, which required only that the IGM have been heated prior to $z \sim 9$.} -- in fact, even colder than that expected in an adiabatically cooling IGM. 

The amplitude of the EDGES signal could be evidence of interactions between baryons and dark matter \citep{Barkana2018}, though a weakly charged DM particle (capable of cooling the baryons through Rutherford scattering) cannot account for the EDGES signal without causing tension elsewhere. For example, \citet{Munoz2018} estimate that if there is a charged DM particle, it can only constitute $\sim 10$\% or less of all of the dark matter (DM). 

Alternatively, \citet{Feng2018} showed that a high-$z$ radio background could supplement the CMB as the illuminating backdrop against which 21-cm absorption is measured, and thus bias the global 21-cm signal toward larger amplitudes. In this work, we will explore both scenarios via flexible additions to our baseline models.

In \citet[][hereafter M17]{Mirocha2017}, we put forth a set of predictions for the global 21-cm signal calibrated to high-$z$ UVLF measurements and with empirically-motivated choices for X-ray source populations. The generic result was a strong absorption trough at high frequencies, $\nu \sim 100 \pm 10$ MHz. Some models of this variety were quickly disfavored, both by EDGES \citep{Monsalve2017} and SARAS \citep{Singh2017}, though most remained viable until the recent report from EDGES \citep{Bowman2018}. A $\sim 78$ MHz trough, regardless of its amplitude, cannot be accommodated by these models in their current form. 

\defcitealias{Mirocha2017}{M17}

Because the \citetalias{Mirocha2017} models were constructed assuming current knowledge of stellar populations and physically-motivated extrapolations of the LF, deviations from our predictions suggest that the UVLF-based view of high-$z$ galaxy evolution is incomplete, regardless of any exotic processes that amplify the signal's strength. A natural first supposition is that very faint sources, like Population III stars and their remnants, could drive the signal to lower frequencies. However, such sources only qualitatively modify the $\sim 100$ MHz expectations under extreme circumstances \citep{Mebane2018,Mirocha2018}. Given the tension between these predictions and the recent EDGES measurement, our goal in this work is to determine what aspects of the astrophysical model -- which is representative of the kinds of models used broadly in the high-z galaxy evolution community \citep[e.g.,][]{Dayal2014,Behroozi2015,Mason2015,Sun2016,Mashian2016} -- require revision. 

In order to test the robustness of our conclusions about high-$z$ galaxies, we explore both signal-amplifying mechanisms \citep[i.e.,][]{Barkana2018,Feng2018} using flexible parametric models. That is, rather than predicting the source(s) responsible for the stronger-than-expected EDGES signal from physical arguments, we attempt to infer the thermal history and radio background that are required to explain the data. Reassuringly, we find that our inferences about galaxy evolution are largely insensitive to the method by which one amplifies the signal's strength. However, the radio background approach places extreme requirements on low-frequency photon production in star-forming galaxies, as we will discuss in \S\ref{sec:results}.

We introduce the 21-cm signal and review our galaxy evolution modeling procedure in Section \ref{sec:methods}, and present results, discussion, and conclusions in Sections \ref{sec:results}, \ref{sec:discussion}, and \ref{sec:conclusions}, respectively. We adopt \citet{Planck2015} cosmological parameters throughout.

\section{Methods} \label{sec:methods}
In this section, we briefly review the global 21-cm signal (\S\ref{sec:gs}), the main features of our galaxy evolution model (\S\ref{sec:lf}), and two possible solutions to the signal's anomalous amplitude (\S\ref{sec:cooling}). The core components of the model (\S\ref{sec:gs}-\S\ref{sec:lf}) are simple extensions of those described in \S2 of \citetalias{Mirocha2017}, so readers familiar with this model may skip ahead to \S\ref{sec:cooling}.

\subsection{The Global 21-cm Signal} \label{sec:gs}
We employ a simple model for the global 21-cm signal in which the IGM is partitioned into two phases \citep[as in, e.g.,][]{Furlanetto2006,Pritchard2010a}. The first is a fully ionized phase, whose sole characteristic is its volume-filling factor, $\QHII$. The second component of the IGM is often referred to as the ``bulk'' IGM, i.e., the IGM outside of ionized bubbles, whose temperature, $\TK$, electron fraction, $x_e$, and $\Lya$ intensity, $\Ja$, must all be considered, as they govern the excitation (or ``spin'') temperature of $\HI$, $\TS$. The global 21-cm signal is simply the volume-averaged brightness temperature of the bulk IGM, down-weighted by the fraction of the volume that is ionized, measured relative to the radiation background temperature\footnote{The most commonly used expression for this so-called ``differential brightness temperature'' is a just solution to the one-dimensional radiative transfer problem, having replaced the usual intensities with temperatures, i.e., adopting the Rayleigh-Jeans approximation).}, usually the CMB, i.e.,
\begin{equation}
    \dTb \simeq 27 (1 - \xibar) \left(\frac{\Obnow h^2}{0.023} \right) \left(\frac{0.15}{\Omnow h^2} \frac{1 + z}{10} \right)^{1/2} \left(1 - \frac{\Tcmb}{T_{\mathrm{S}}} \right) , \label{eq:dTb}
\end{equation}
where $\xibar = \QHII + (1 - \QHII) x_e$ is the volume-averaged ionized fraction, and
\begin{equation}
    T_S^{-1} \approx \frac{T_{\gamma}^{-1} + x_c T_K^{-1} + x_{\alpha} T_{\alpha}^{-1}}{1 + x_c + x_{\alpha}} . \label{eq:Ts}
\end{equation}
In words, Equation \ref{eq:Ts} means that the spin temperature of $\HI$ is set by collisions, whose coupling strength depends on the coefficient $x_c$ \citep[which itself depends on density, temperature, and ionization state;][]{Zygelman2005} and the kinetic temperature (determined by the balance between heating from sources and cooling due to cosmic expansion), and radiation backgrounds, including the CMB, with temperature $\Tcmb$, and the $\Lya$ background, characterized by $T_{\alpha}$. The latter dependence is not obvious -- spontaneous absorption and re-emission of $\Lya$ photons can result in a spin-flip \citep[see, e.g., Fig. 1 of][]{Pritchard2006}, a byproduct of quantum selection rules for the total spin angular momentum. This is known as the Wouthuysen-Field effect \citep{Wouthuysen1952,Field1958}, with a magnitude quantified by the radiative coupling coefficient is $x_{\alpha} = 1.81 \times 10^{11} \widehat{J}_{\alpha} S_{\alpha} / (1 + z)$, where $J_{\alpha}$ is the background $\Lya$ intensity. The factor $S_{\alpha}$ accounts for line profile effects \citep{Chen2004,FurlanettoPritchard2006,Chuzhoy2006,Hirata2006}. The high optical depth of the $\Lya$ line rapidly equilibrates the radiation and kinetic temperatures such that in practice, $T_{\alpha} \simeq T_K$.

The time evolution of the IGM's properties, $\QHII$, $\TK$, and $x_e$, is of course governed by the properties of sources in the volume. We construct a model for the volume-averaged emissivity of sources as a function of time (to be discussed in the next sub-section), and evolve the emergent radiation field using standard techniques \citep[as in, e.g.,][]{Haardt1996} to obtain the mean meta-galactic background intensity, $J_{\nu}$, at all redshifts. With $J_{\nu}$ in hand, ionization and heating rates can be calculated, and the state of the gas in each phase of our two-zone IGM can be evolved in time. We perform these computations with the \ares\ code\footnote{\url{https://bitbucket.org/mirochaj/ares}}, which has been described in greater detail elsewhere \citep{Mirocha2014}.

\subsection{High-$z$ Galaxies} \label{sec:lf}
We adopt a simple model for high-$z$ galaxy evolution, which assumes that star formation is fueled by the smooth inflow of pristine gas from the IGM into galaxies, i.e., $\SFR = \fstar \MAR$. Similar techniques have been employed by several groups in recent years, e.g., \citet{Behroozi2015,Mason2015,Mashian2016,Sun2016}. Whereas some models using merger trees to construct halo growth trajectories, we adopt a simpler approach. The growth rate of halos, $\MAR$, is computed assuming halos grow at fixed number density. Though an over-simplification, this yields fair agreement with the results of numerical simulations \citep[e.g.,][]{McBride2009} where they overlap (at $3 \lesssim z \lesssim 6$), which find roughly that $\dot{M}_h \propto M (1+z)^{5/2}$. This approach is convenient also because it guarantees self-consistency with the adopted halo mass function\footnote{We assume a \citet{ShethMoTormen2001} form, which we generate using the \textsc{hmf} code \citep{Murray2013}.} \citep[see][Appendix A]{Furlanetto2017}.

With a model for the abundance of galaxies and their growth rates, one can empirically calibrate the star formation efficiency (SFE), $\fstar$, by fitting the model to UVLF measurements. We assume that the luminosity of galaxies at all wavelengths is dominated by star formation, such that $L_{h,\nu} = \SFR(M_h, z) l_{\nu}$, where $l_{\nu}$ is the specific luminosity per unit SFR. Then, assuming a 1:1 correspondence between DM halos and galaxies, we can derive the UVLF (at the usual rest-frame $1600$ \AA) as
\begin{equation}
    d\phi(L_h) = \frac{d n(M_h, z)}{dM_h} \left( \frac{dL_h}{d M_h} \right)^{-1} d L_h . \label{eq:LF}
\end{equation}
where $n(M_h, z)$ is the number density of halos of mass $M_h$ at redshift $z$, and $\phi$ is the number density of galaxies with luminosity $L_h$. 

In \citetalias{Mirocha2017}, we calibrated to the \citet{Bouwens2015} UVLF at $z=5.9$ assuming a double power-law form for the SFE. Though a redshift-independent SFE is known to result in galaxy populations broadly consistent with current measurements over $0 \lesssim z \lesssim 8$ \citep[e.g.,][]{Behroozi2013,Mason2015}, application of these models to very high-$z$ cannot explain the EDGES measurement, as we will show in the next section. As a result, we allow the normalization, peak mass, and slopes of the SFE (at masses above and below the peak) to evolve with redshift as power laws. In addition, we allow the SFE to deviate from a single power-law at the faint-end, either by reaching a floor, or declining exponentially to zero below some critical mass. These phenomenological extensions are allowed to vary with redshift as well. We do not attempt to model scatter in the luminosity of galaxies (at fixed mass), caused either by scatter in the dust correction or star formation rates, as these effects mostly affect the bright-end of the luminosity function, to which the global 21-cm signal is least sensitive.

The volume-averaged emissivity, which seeds the meta-galactic radiation background, can be computed as a weighted integral over the luminosity function assuming that radiation at all wavelengths is dominated by star formation. We use the BPASS version 1.0 models \citep{Eldridge2009} to generate the UV luminosities of galaxies assuming continuous star formation, and adopt an empirically-calibrated relation between star formation rate (SFR) and X-ray luminosity, $L_X$, to synthesize the X-ray background. Our fiducial model assumes the $L_X$-SFR relation from \citet{Mineo2012a}, who find that HMXBs are the dominant sources of X-ray emission in local star-forming galaxies, emitting $L_X = 2.6 \times 10^{39} \ \cXunits$ in the 0.5-8 keV band. We assume an unabsorbed multi-color disk spectrum \citep{Mitsuda1984} extending from 0.2 keV to 30 keV for our calculations. Order of magnitude boosts in the $L_X$-SFR relation may be possible at high-$z$ in the low metallicity environments \citep{Brorby2016} of the first galaxies, so we allow the $L_X$-SFR normalization to scale by a factor $f_X$, left as a free parameter in our fits. 

In this work, we assume that only galaxies in halos above the atomic cooling threshold, with virial temperatures $\Tmin \gtrsim 10^4$ K, contribute to the volume-averaged emissivity. Our goal is to determine if the known galaxy population, with some extrapolation to magnitudes and redshifts beyond current detection thresholds, can account for the EDGES signal. Entirely ``new'' sources (i.e., those that have yet to be observed directly), of which PopIII stars and their remnants are plausible candidates, could of course also impact the signal. However, PopIII stars should have a rather subtle impact on the global 21-cm signal in all but the most extreme cases \citep{Mirocha2018}, hence our focus on more massive galaxies for the time being. There could be some ambiguity in the interpretation of our results, even if PopIII stars are not ultimately important. For example, are the extrapolations we find necessary to accommodate the EDGES signal indicative of real changes in how galaxies function at the highest redshifts? Or, are they symptoms of problems with the model itself, or neglect of other, more familiar source populations (e.g., globular clusters, AGNs)? We defer a detailed discussion of this issue to \S\ref{sec:discussion}. 

Finally, before moving on, we emphasize that Equation \ref{eq:LF} is equivalent to the \textit{observed} LF only under the assumption that all photons in the observed band (rest-frame $1600$ \AA\ here) escape galaxies. In general, this is not the case, as some rest-frame $1600$ \AA\ photons will be absorbed by dust before they can escape the galaxy. As a result, we extend the \citetalias{Mirocha2017} calibration to the entire redshift range $4 \lesssim z \lesssim 10$ using a \citet{Meurer1999} relation between extinction, $A_{\mathrm{UV}}$, and UV slope, $\beta$, inferring $\beta$ through the $\beta$-$M_{\mathrm{UV}}$ relation of \citep{Bouwens2014}. This is in contrast to \citetalias{Mirocha2017}, in which we neglected dust, in part motivated by the seemingly-low dust content in some high-$z$ galaxies \citep{Capak2015}. Including dust has only a minor effect on the global signal -- changing the star formation rate density (SFRD) by a factor of $\sim 2$ -- though the impact on UVLFs is large, especially at the bright end. Rising dust temperatures could be responsible for the observed dust deficit \citep[e.g.,][]{Narayanan2018}, though we make no attempt to model the time evolution of the dust contents of galaxies in this work \citep[as in, e.g.,][]{Imara2018}.

\subsection{Mechanisms for Amplifying the Differential Brightness Temperature} \label{sec:cooling}
In order to even qualitatively match the EDGES measurement our models require a process capable of increasing the amplitude of the global 21-cm absorption signal beyond that which is expected in an adiabatically cooling IGM. This can be accomplished via new coolants, such as charged DM \citep{Barkana2018}, which can reduce the gas temperature below that predicted in standard $\Lambda$CDM models of recombination \citep[e.g., using \textsc{cosmorec};][]{Chluba2011}. Alternatively, any radio backgrounds present with an intensity comparable to the CMB would supplement $\Tcmb$ in Equation \ref{eq:dTb} and cause stronger absorption signals (at fixed $\TS$). Indeed, such backgrounds could amplify the global 21-cm signal substantially without causing tension elsewhere \citep{Feng2018}, as there is an excess in the observed cosmic radio background at frequencies $\nu \lesssim 1$ GHz of order $\sim$few Kelvin, rapidly rising to $\gtrsim 10^3$ K at $\nu \lesssim 100$ MHz \citep{Fixsen2011}. The observed excess could be a result of instrumental systematics or new sources \citep[see conference summary by ][]{Singal2018}, hence the need to explore the plausibility of new sources at high-$z$ in light of the strong EDGES signal.

We explore the effects of both mechanisms for global 21-cm signal amplification, which we describe briefly below. 

\subsubsection{A Parametric Approach to ``Excess'' Cooling} \label{sec:excess_cooling}
\citet{Barkana2018} found that milli-charged DM can provide an additional cooling channel for the baryons so long as its mass is $m_{\chi} < 23$ GeV and its cross section is $\sigma > 3.4 \times 10^{21} \ \mathrm{cm}^{-2}$. However, \citet{Munoz2018} have argued that this type of DM cannot constitute the entirety of DM without violating constraints on the local DM density. Given that the origin of the excess cooling is still up for debate, we take a model-agnostic view. Rather than appealing to a particular physical model for DM, we parameterize the thermal history using a form that can (i) accurately recover a case in which the thermal history proceeds ``normally,'' i.e., as recombination codes predict \citep[e.g., \textsc{cosmorec}][]{Chluba2011}, and (ii) allow cooling to occur more rapidly and/or at earlier times than in typical models. 

To do this, we recognize that at high-$z$ the temperature of a mean-density gas parcel evolves between $T(z) \propto (1+z)$ and $T(z)=(1+z)^2$ -- the $(1+z)$ dependence a signature that Compton scattering tightly couples the CMB, spin, and kinetic temperatures, and the $(1+z)^2$ dependence indicating that Compton scattering has become inefficient, allowing the gas to cool adiabatically. The simplest parameterization of the thermal history thus appears to be a broken power-law in redshift. However, such a thermal history can also be recovered by noting that the \textit{log}-cooling rate, $d\log T/ d\log t$, of a mean-density gas parcel transitions smoothly between $-2/3$ at very high-$z$ and $-4/3$ in a matter-dominated cosmology. So, rather than modeling the thermal history directly, we take
\begin{equation}
    \frac{d\log T}{d\log t} = \frac{\alpha}{3} - \frac{(2+\alpha)}{3} \bigg\{1 + \exp \left[-\left(\frac{z}{z_0}\right)^{\beta} \right] \bigg \} \label{eq:Thist}
\end{equation}
and integrate to obtain the thermal history. We have constructed this relation such that $\alpha=-4$ reproduces the typical thermal history, and while varying $\alpha$ can change the late-time cooling rate, the cooling rate as $z \rightarrow \infty$ tends to $d\log T/ d\log t = -2/3$, as it must to preserve the thermal history during the recombination epoch. We find that this form does a better job recovering thermal histories from \textsc{Cosmorec} than does a broken power-law for $T(z)$. It also outperforms a hyperbolic tanh function for the log cooling rate, which is another natural choice when modeling smooth transitions between known limits. Setting $z_0=189.6$, $\beta=1.27$, and $\alpha=-4$ reproduce the thermal history and its derivative generated with \textsc{cosmorec} (assuming \citet{Planck2015} cosmological parameters) to better than 1\%. The parameter $z_0$ is a decoupling redshift, indicating the redshift at which the cooling rate is halfway between the asymptotic limits, while $\beta$ indicates how rapidly the cooling rate declines after Compton scattering becomes inefficient. Larger values of $\beta$ indicate sharper declines in $d\log\TK/d\log t$. There is of course a degeneracy between these parameters, but, for the purposes of interpreting the $z \sim 18$ EDGES measurement, the detailed values of $z_0$ and $\beta$ are unimportant. Measuring the underlying cooling rate will require observations at $\nu \lesssim 20$ MHz, where modifications to standard thermal histories can be studied without the complicating influence of astrophysical sources \citep[see Fig. 1 of ][]{Fialkov2018}. Such observations will likely require observations from space \citep[e.g., DARE][]{Burns2017}.

We allow astrophysical sources to begin forming at $z=60$, at which point we solve the thermal history by numerically integrating the standard evolution equations, which depend on the rate of heating (from sources) and cooling (from Equation \ref{eq:Thist}). Note that while we do not self-consistently solve for the very high-$z$ thermal and ionization histories, any modifications to the post-recombination electron fraction have very little impact on the 21-cm background, since H-$e^{-}$ collisions are sub-dominant to H-H collisions in setting the spin temperature unless the electron fraction is order unity.

\subsubsection{A Radio Background Excess} \label{sec:radio_excess}
The global 21-cm signal may also be amplified relative to common expectations if the CMB is not the only radio background at very high-$z$. \citet{Feng2018} have worked backwards from the ARCADE-2 excess \citep{Fixsen2011}, and found that stronger-than-expected global 21-cm signals are possible even if only $\sim 10$\% of the $z=0$ excess is from high-$z$. However, they did not attempt to model the source of such a high-$z$ background.

We investigate the possibility that star-forming galaxies themselves source the radio background, and thus account for all features of the observed global 21-cm signal. There are known relations between SFR and radio luminosity \citep[e.g.,][]{Condon2002,Heesen2014,Gurkan2018}, which we can easily implement in our models and augment $\Tcmb$ accordingly at all redshifts. As our reference point, we adopt an empirical relation between the monochromatic 150 MHz luminosity and SFR \citep[as in, e.g.,][]{Gurkan2018},
\begin{equation}
    L_R = 10^{22} f_R \left(\frac{\mathrm{SFR}}{\Msun \ \mathrm{yr}^{-1}} \right) \ \mathrm{W} \ \mathrm{s}^{-1} \ \mathrm{Hz}^{-1} \label{eq:LrSFR}
\end{equation}
and extrapolate to higher frequency assuming a spectral index of $-0.7$, as in \citet{Gurkan2018}.

We treat $f_R$ as a free parameter, and aim to quantify what its value must be in order to explain the EDGES measurement. Also, we will show in the next section that truncating the radio background at some critical redshift, $\zoff$, is necessary in order to match both the shape of the EDGES signal and to satisfy the $z=0$ ARCADE-2 excess. To our knowledge, the possibility of redshift evolution in the $L_R$-SFR relation has not been explored, so we hope our results may help to motivate and contextualize future theoretical models. 

\section{Results} \label{sec:results}

\subsection{Qualitative Expectations}
The absorption peak detection by EDGES is centered at $\nu \sim 78$ MHz, some $\sim 10-30$ MHZ lower in frequency than the reference models from \citetalias{Mirocha2017}. In other words, Wouthuysen-Field coupling and X-ray heating occurred $\sim 100-300$ Myr earlier in time than simple UVLF-based schemes suggest. If we are to explain the EDGES signal with the ``normal'' galaxy population, we must amplify the total amount of star formation occurring at $z \gtrsim 10$, the amount of $\Lya$ and X-ray emission \textit{per unit} star formation at $z > 10$, or both, relative to the baseline \citetalias{Mirocha2017} model \citep[similar to other UVLF-based models;][]{Sun2016,Mason2015,Mashian2016}. This is curious because, based on UVLFs alone, most theoretical models of galaxy evolution actually \textit{over}-produce UV-bright galaxies at $z \sim 10$ \citep{Oesch2017}.

\begin{figure*}
\begin{center}
\includegraphics[width=0.98\textwidth]{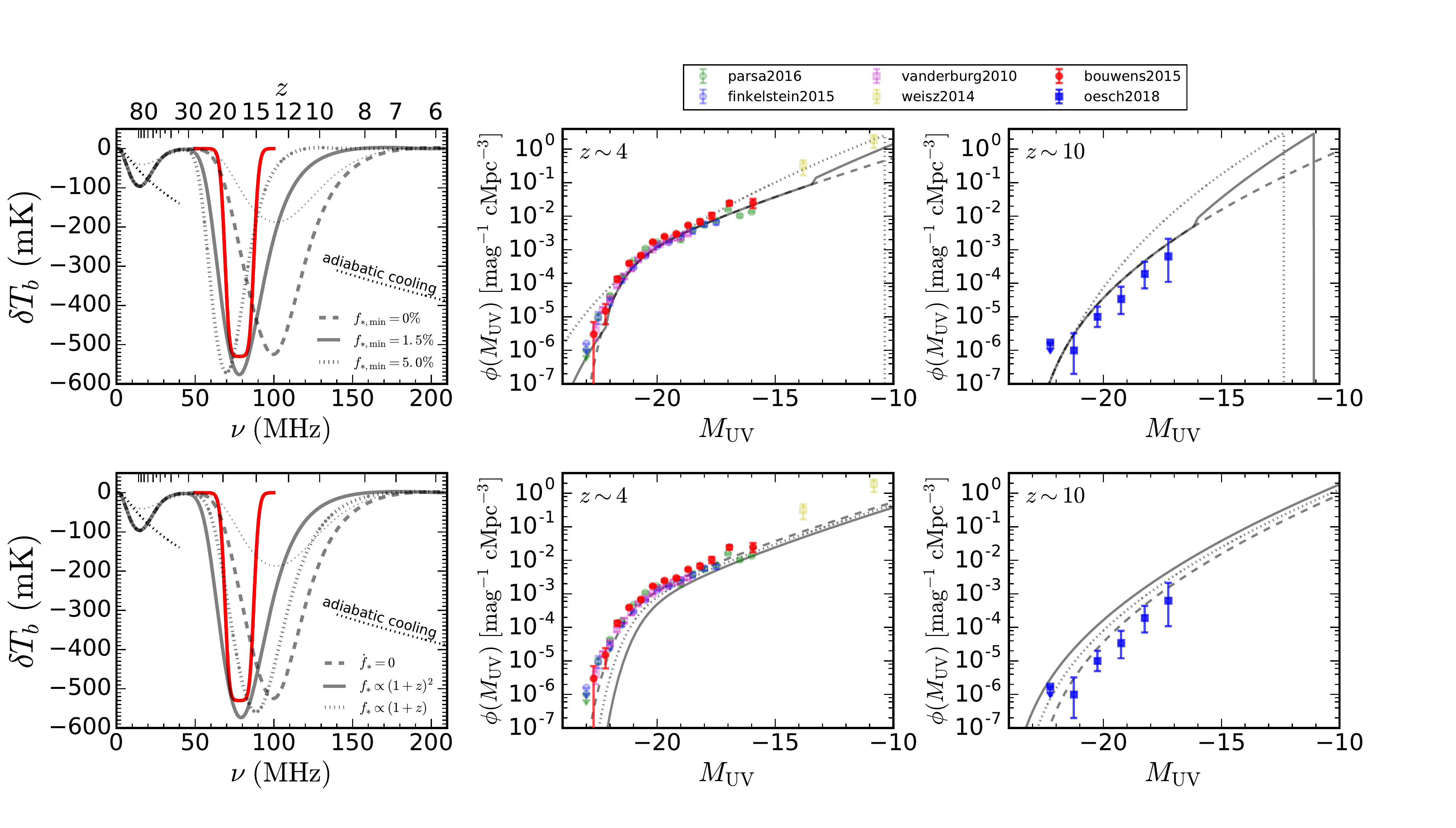}
\caption{{\bf Competing requirements of UVLFs and the EDGES signal.} In general, driving the 21-cm absorption trough requires the addition of ``new'' star formation -- the top row appeals to a change in the shape of the SFE in low-mass objects (by introducing a floor in the SFE), while the bottom row instead invokes time evolution in the overall normalization of the SFE. \textit{Left:} Both methods of boosting star formation at high-$z$ can recover 21-cm absorption troughs at $\sim 80$ MHz, though the modifications required to do so are substantial, i.e., a 1.5\% floor in the SFE (top left) or $(1+z)^2$ evolution in its normalization (bottom left). However, our example SFE modifications manifest in the UVLF as well, as shown at $z \sim 4$ (\textit{middle column}) and $z \sim 10$ (\textit{right column}). Enhanced star formation in low-mass objects steepens the faint-end of the UVLF, which, if too aggressive, causes tension with directly-measured high-$z$ UVLFs. However, the UVLF reconstructions from \citet{Weisz2014} at $z \sim 4$ prefer a steepening. Redshift evolution in the SFE normalization (bottom) leads to considerable tension with both the $z\sim 4$ and $z \sim 10$ LF constraints, for models that match at $z\sim 6$ by construction, assuming the normalization evolves monotonically in time as a power-law (indicated in legend). Note that if $z \sim 10$ galaxies have systematically low dust contents, the disagreement at $z \sim 10$ will become even worse.}
\label{fig:param_study}
\end{center}
\end{figure*}

This problem is shown graphically in Figure \ref{fig:param_study}. Driven by the need to match the amplitude and frequency of the EDGES signal\footnote{None of our models match the EDGES measurement's shape in detail. We will revisit this in \S\ref{sec:discussion}.}, we show two primary means by which to increase the star formation rate density at high-$z$: introducing a floor in the SFE, which amplifies star formation in faint galaxies only (top), and invoking a rise in the normalization of the SFE with redshift (bottom). We show the global signal results in the left-most column, with the consequences for UVLFs in the center and right columns, compared to a highly incomplete set of results from the recent literature\footnote{We have made no effort to homogenize these data, some of which correspond to slightly different redshifts (at the level of $\Delta z \sim 0.1$) and rest-wavelengths ($\Delta \lambda \sim 100-200$ \AA).} \citep{Vanderburg2010,Parsa2016,Weisz2014,Bouwens2015,Finkelstein2015,Oesch2017}. Note that the \citet{Weisz2014} measurements are unlike the rest, in that they are the \textit{reconstructed} UVLFs of local dwarfs, i.e., projections of their star formation histories back in time. In each panel, dashed curves denote the predictions of the reference \citetalias{Mirocha2017} model, solid black curves indicate models that roughly match the EDGES detection, while dotted curves merely provide another example to indicate sensitivity to the parameters.

Starting in the bottom row, we see that the $\fstar \propto (1+z)^2$ dependence required to match the peak of the EDGES signal (solid curves) quickly generates tension with observed UVLFs within 500 Myr of the end of reionization, both at $z \sim 4$ and $z \sim 10$ (having anchored the model to $z \sim 6$ UVLFs). Because we have assumed monotonic redshift evolution, this model over-predicts the abundance of galaxies at $z \sim 10$ and underestimates the abundance of galaxies at $z \sim 4$. A model with $\fstar \propto (1+z)$, as is expected from some analytic feedback models \citep{Furlanetto2017}, still over-produces $z \sim 10$ galaxies and under-produces $z \sim 4$ galaxies, while remaining in tension with the EDGES measurement by $\sim 10$ MHz. 

In the top row, we see that introducing a floor in the SFE can reduce tension between the UVLFs and EDGES ``for free,'' i.e., without violating any observational constraints. In fact, the reconstructed star formation histories of local dwarfs seem to suggest a steepening in the faint-end of the UVLF at these redshifts \citep{Weisz2014}, as shown in the top-middle panel. The dotted curve, which imposes a floor in the SFE at 5\% (independent of halo mass) matches the \citet{Weisz2014} points, but generates a trough in the global 21-cm slightly too early, and departs from the $z\sim 10$ UVLF fairly substantially.

Because the highest redshift UVLFs contain only the brightest galaxies, one solution to the problem at hand is to simultaneously reduce the SFE of high-mass halos with $z$ while amplifying the SFE of low-mass halos. This implies a change in both the shape and the normalization of the SFE with time, which, as we will see shortly, tends toward a solution in which galaxy star formation rates scale linearly with mass, i.e., the star formation efficiency becomes independent of mass. We will postpone a physical interpretation of this result to \S\ref{sec:discussion}.

\subsection{Fitting the EDGES Signal}
Due to the complex interplay between factors discussed in the previous sub-section, we perform a multi-dimensional fit \citep[using \textsc{emcee};][]{ForemanMackey2013} to explore a broad range of possibilities. We fit the \citet{Bouwens2015} UVLFs from $4 \lesssim z \lesssim 8$, and the $z\sim 10$ UVLF presented in \citet{Oesch2017}. Use of other UVLFs can systematically shift the inferred SFE, though the effect is minor \citep[see, e.g.,][]{Mason2015}. For the global 21-cm measurement, we adopt the recovered signal shown in Figure 1 of \citet{Bowman2018}, with peak amplitude $T_{21}=-530$ mK, central frequency $\nu_0 = 78.1$ MHz, width of 18.7 MHz, and flattening factor $\tau=7$. We adopt a frequency-independent uncertainty of 100 mK (conservative root-mean squared residual after removing the foreground) across the EDGES band, deferring a more detailed treatment of covariances between model parameters, instrument, and foreground to future work. 

We vary the parameters that control the SFE, including the normalization (defined as $\fstar$ at $M_h = 10^{10} \ M_{\odot}$), peak mass, low- and high-mass slope, and faint-end modifications (floor or steep decline), allowing each quantity to evolve with redshift. Furthermore, we allow $\Tmin$ and $f_X$ to vary in the fit. This results in 16 total parameters, the majority of which (14/16) are in place to describe the UVLF from $4 \lesssim z \lesssim 10$. For the excess cooling model, we also vary $\alpha$, $\beta$, and $z_0$ (see \S\ref{sec:excess_cooling}), and for the radio background model we add $f_R$ to the list of free parameters (see \S\ref{sec:radio_excess}).

We do not vary the stellar metallicity\footnote{For simplicity, we adopt solar metallicity ($Z=0.02$), which means our inferred star formation rates can be scaled by the ratio of 1600 \AA\ luminosities of stellar populations of different metallicities. However, this is only a $\sim$ 10\% effect (between $Z = 0.001$ and $Z=0.02$) in the BPASS v1.0 models \citep{Eldridge2009}, for example, so this hardly affects our results.}, as it is completely degenerate with the SFE in the context of our model \citepalias[see \S3.4 in][]{Mirocha2017}, nor do we vary the escape fraction, as the IGM must be nearly neutral at the redshifts of the EDGES signal, so the new detection offers no direct information on that parameter. We will comment on the implications for the reionization history in \S\ref{sec:discussion}.

\begin{figure*}
\begin{center}
\includegraphics[width=0.98\textwidth]{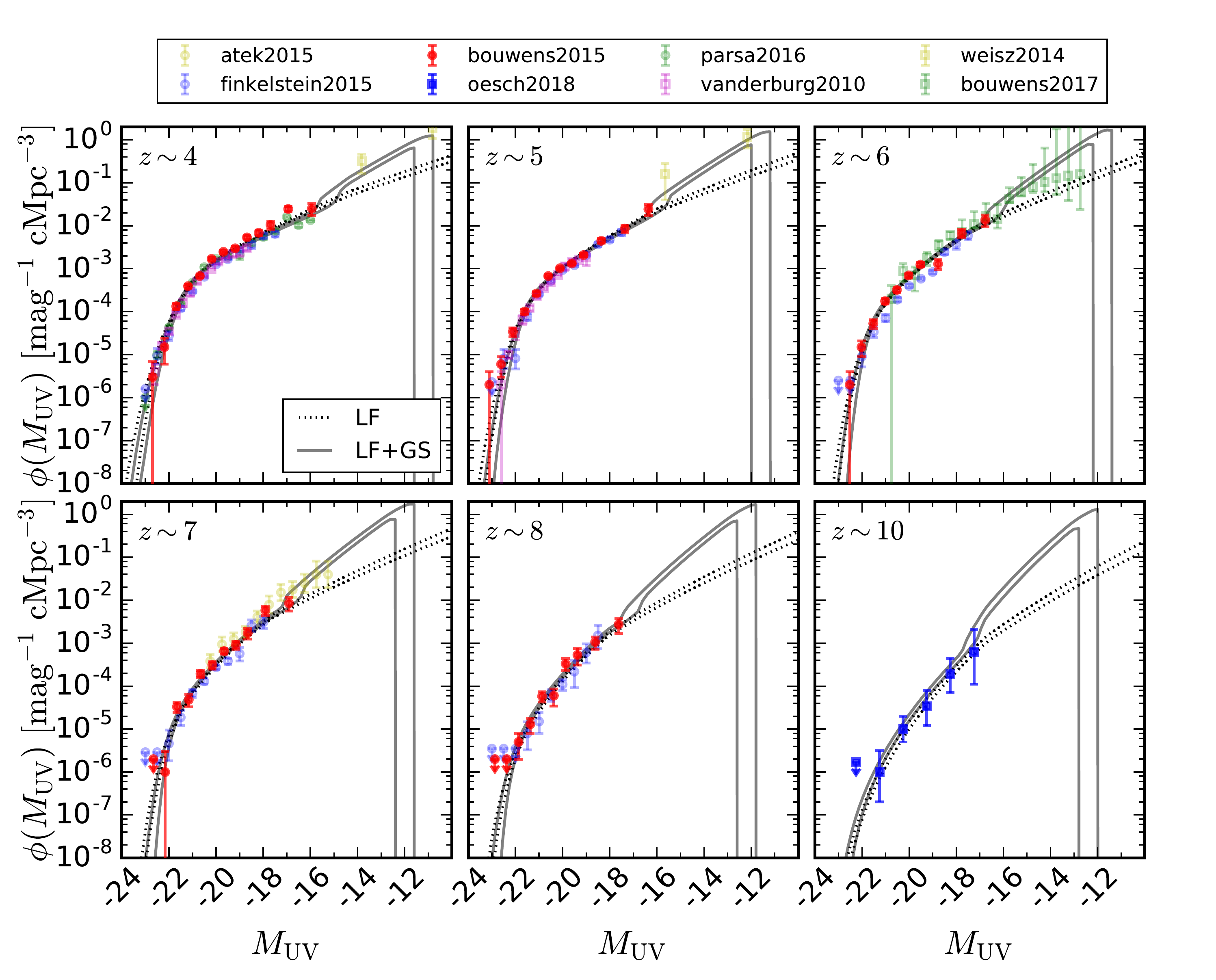}
\caption{{\bf Recovered galaxy luminosity functions, with (solid) and without (dotted) knowledge of the global 21-cm signal.} The EDGES signal demands much stronger early star formation than anticipated from an extrapolation of the observed UVLFs, which manifests in a substantial steepening of the LF in the (so far unobserved) faint regime. The pair of lines for each model bracket the 68\% confidence intervals in each fit. Because we have no mechanism in place to suppress galaxy formation in low-mass halos (e.g., reionization feedback), the enhanced star formation in low-mass objects required to fit the EDGES measurement persists to arbitrarily low redshifts. Note that for simplicity, only the filled plot symbols have been used in the fit \citep[i.e.,][]{Bouwens2015,Oesch2017}. We have also added points from \citet{Atek2015} and \citet{Bouwens2017}, which represent some of the deepest limits yet, made possible by the Hubble Frontier Fields program \citep{Lotz2017}.}
\label{fig:lf_recon}
\end{center}
\end{figure*}

\subsection{Implications for Galaxy Formation Scenarios}

In Figure \ref{fig:lf_recon}, we show a compilation of UVLF measurements from the literature along with the results of two fits. 

First, we present our reference model fits to UVLFs drawn from \citet{Bouwens2015} and \citet{Oesch2017} at $4 \lesssim z \lesssim 10$ as dotted lines, neglecting the EDGES measurement entirely. It is an update of the \citetalias{Mirocha2017} reference model in that it corrects UVLFs for dust attenuation (see \S\ref{sec:lf}) and simultaneously fits data over a broad range of redshifts\footnote{The dust correction results in higher a inferred value for the SFE in the brightest objects, at the factor of $\lesssim 2$ level.}. We also allow the normalization of the SFE to evolve in time as a power-law, as is expected in some theoretical models \citep[e.g.,][]{Furlanetto2017}.

Second, in the solid lines we show the results when we simultaneously fit UVLFs and the EDGES measurement, assuming Equation \ref{eq:Thist} for the cooling rate of the high-$z$ IGM. Inclusion of the EDGES measurement requires a boost in the star formation efficiency in the faintest galaxies, as evidenced by the steepening faint-end slope. Such faint objects have thus far only been found in lensing fields, in which their abundance remains controversial \citep{Livermore2017,Bouwens2017}. However, the  star-formation histories of local dwarfs suggest that such luminosities may not be unreasonable \citep{Weisz2014}. These slopes correspond to a flat star formation efficiency, and will thus roughly mirror the slope of the DM halo mass function. Note that fitting the EDGES measurement of course does not imply such faint-end behavior is required at $z \sim 4$. This is a byproduct of only allowing the SFE to evolve monotonically in time -- in reality, reionization feedback could stifle star formation in low-mass objects at late times, even if such objects indeed harbor unexpectedly efficient star formation at $z \sim 18$. 

\begin{figure*}
\begin{center}
\includegraphics[width=0.98\textwidth]{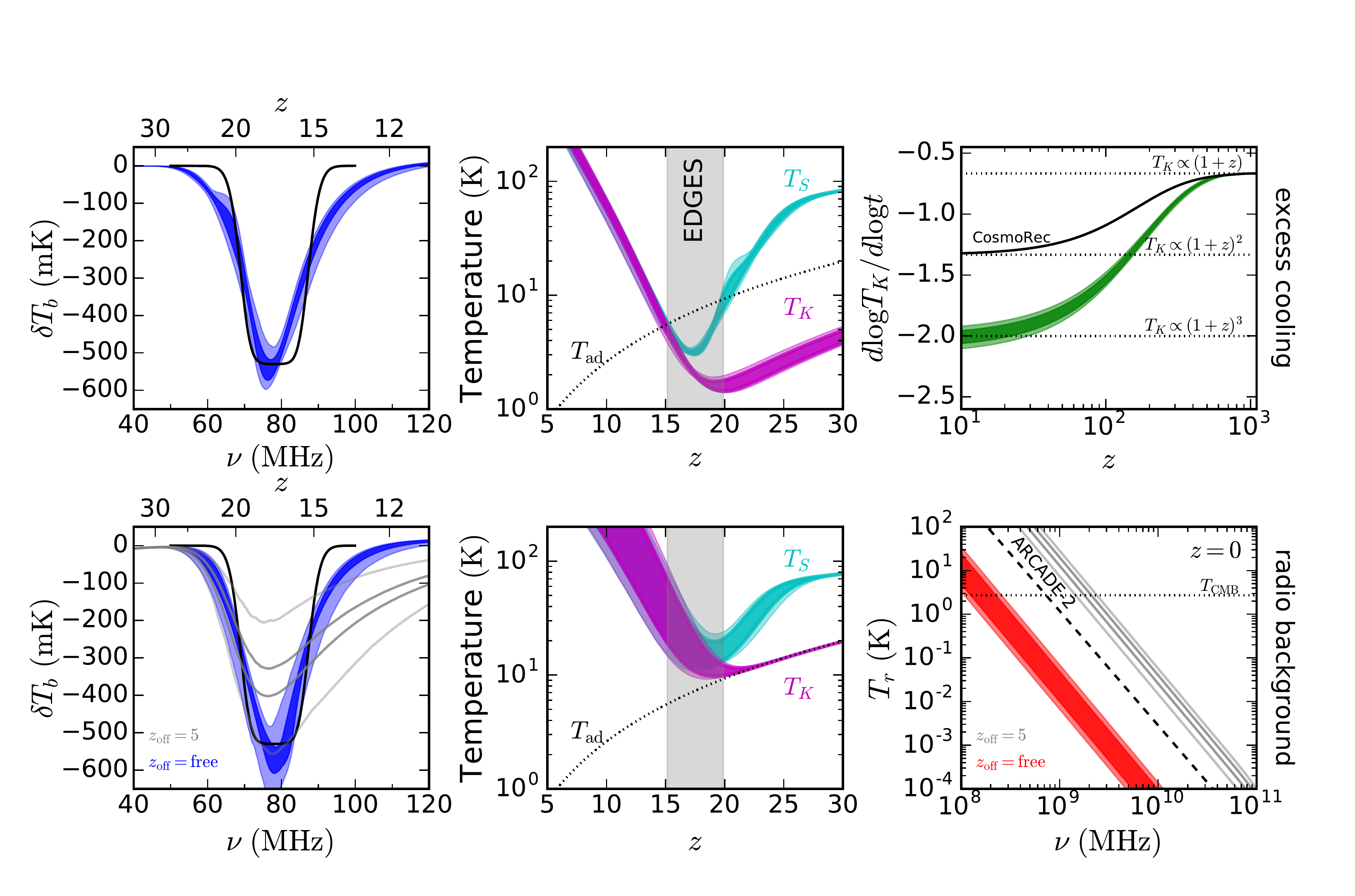}
\caption{{\bf Reconstructed global 21-cm signal, thermal history, and radio background.} \textit{Top:} Results for an excess cooling model, in which the IGM cools following Equation \ref{eq:Thist}. From left to right, we show the 68\% and 95\% confidence interval for reconstructed global 21-cm signal, thermal history, and cooling rate evolution. For reference, we also show the adiabatic limit (dotted black; middle panel), which corresponds to $d\log \TK / d\log t = -4/3$ (or $\TK \propto (1+z)^2$; right panel). \textit{Bottom:} Results for a radio background model, in which $L_R \propto f_R \SFR$. Again, we show the global signal and thermal history in the left-most panels, but now show the $z=0$ radio background compared to the reported ARCADE-2 excess \citep[dashed diagonal line;]{Fixsen2011} and the CMB (dotted horizontal line). Open gray contours indicate the solution obtained if one allows the radio emission to continue until the calculation terminates at $\zoff=5$, while filled contours show the results when $\zoff$ is left as a free parameter. Our best-fit yields $\zoff \sim 15$ and $f_R \sim 10^3$, indicating that a strong -- but quickly truncated -- background is required both to match the shape of the EDGES signal and also remain below the $z=0$ excess (right panel).}
\label{fig:gs_recon}
\end{center}
\end{figure*}

Results for the global signal, thermal history, and radio background are shown in Figure \ref{fig:gs_recon}. Each row adopts a different mechanism for increasing the amplitude of the signal, including a parametric cooling excess (top), and a radio background sourced by galaxies at high-$z$ (bottom). 

Focusing first on the top row, we see that our reconstructed global 21-cm signal, while broadly consistent with the EDGES measurement, cannot match its shape in detail (left panel), as it lacks a flattened peak and has broader wings than the EDGES signal. The temperature evolution implied by this realization is explored in the right two panels. First, though the spin temperature at $z \sim 18$ is $\sim 3-4$ K, the kinetic temperature is only $\sim 1-2$ K, as radiative coupling is not fully complete. In order to achieve such temperatures -- assuming our parametric thermal history is reasonable -- cooling rates must exceed the adiabatic rate at redshifts $z \gtrsim 100$ (right panel). Typically (i.e, in the absence of exotic mechanisms) cooling doesn't become fully adiabatic until $z \lesssim 20$.

In the bottom row of Figure \ref{fig:gs_recon}, we present results obtained assuming the $L_R$-SFR relation of Equation \ref{eq:LrSFR}. Our best-fitting normalization requires $f_R \sim 10^3$, which is a strong requirement of star-forming galaxies. If such a strong $L_r$-SFR relation is allowed to persist through the end of reionization, a $z=0$ radio excess even stronger than that reported by ARCADE-2 is implied (gray contours; right panel). As a result, we also allow the boosted radio emission to terminate at some critical redshift, $\zoff$. Though an ad-hoc modification of the model, interestingly, it alleviates tension at $z=0$ while simultaneously producing a sharper global signal, as shown in the blue contours in the bottom left panel of Figure \ref{fig:gs_recon}. Without such truncation, the radio background grows monotonically with time as the CMB decays, resulting in a signal with a broad tail to high frequencies. The thermal history in these models is consistent with standard models, implying IGM temperatures of $\sim 10-20$ K at $z \sim 18$.

The thermal histories we recover are sensitive to the SFRD and $f_X$. In our excess cooling model, we recover $f_X \sim 10$, while in the radio excess model we can only limit $f_X \gtrsim 10$, as arbitrarily large values of $f_X$ can be counteracted by commensurate boosts in $f_R$. In fact, these values are strong lower limits, given that we have assumed solar metallicity (our SFRD is an overestimate if high-$z$ galaxies are sub-solar), and an unabsorbed X-ray spectrum \citep[neutral absorption can significantly reduce heating at fixed $f_X$;][]{Mirocha2014,Das2017}. Should rapid heating continue throughout reionization, we should expect a strong emission feature in the global 21-cm signal at frequencies $\nu \gtrsim 100$ MHz. So far, there are only lower limits on the duration of reionization ($\Delta z \gtrsim 1$;assuming a saturated 21-cm signal) from the EDGES high-band receiver \citep{Bowman2010,Monsalve2017}.

\begin{figure*}
\begin{center}
\includegraphics[width=0.98\textwidth]{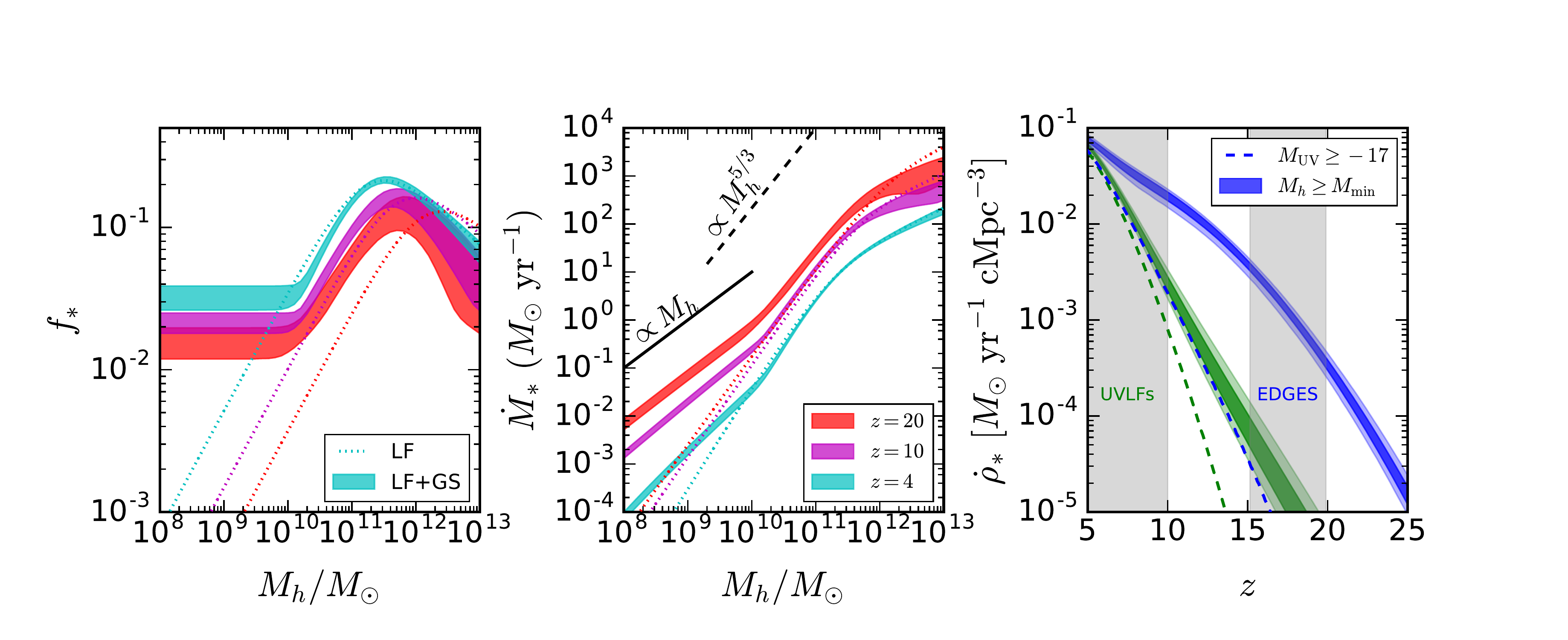}
\caption{{\bf Recovered relationships between DM halos and star formation in galaxies.} Dotted curves correspond to UVLF-only solutions in each panel, while solid (or filled) contours indicate the results obtained when simultaneously fitting UVLFs and \edges. \textit{Left:} Efficiency of star formation, defined as the fraction of inflowing gas converted into stars. \textit{Middle:} Relationship between SFR and halo mass. A flat SFE translates into a linear curve in this space, since DM growth rates scale roughly at $\dot{M}_h \propto M_h$. \textit{Right:} Reconstructed SFRD using UVLFs only in the fit (green) vs. UVLFs and the EDGES signal (blue). Dashed curves indicate the best-fitting SFRD integrated only above $M_{\mathrm{UV}} \geq -17$ for each case, while the filled contours show 68\% and 95\% confidence intervals for the total SFRD, i.e., integrating down to the minimum mass threshold.}
\label{fig:sfrd_recon}
\end{center}
\end{figure*}

In Figure \ref{fig:sfrd_recon}, we show the reconstructed relationships between dark matter halos and star formation with time, which give rise to changes in the UVLF and global signal shown in Figures \ref{fig:lf_recon} and \ref{fig:gs_recon}. 

Starting in the left panel, we show the SFE at three redshifts, comparing the full solution obtained upon fitting UVLFs and the EDGES signal (bands), as well as the standard approach with UVLFs only (dotted lines). The decline in the normalization of the SFE with time is driven by the UVLF alone (mostly the $z \sim 8$ and $z \sim 10$ points), and scales roughly as $\fstar \propto (1+z)^{-1}$ in $M_h = 10^{10} \ \Msun$ halos. Though the fit could in principle have generated a model in which the slope of the SFE in low-mass objects gradually evolved in time, the actual recovered curves show a preferred departure from the power-law below $M_h \sim 10^{10} \ \Msun$. This is the source of the steepening in UVLFs shown in Figure \ref{fig:lf_recon}. A mass of $10^{10} \ \Msun$ corresponds roughly to current sensitivity limits, which, coupled with the flat SFE we infer, means the optimal fit to the UVLF and EDGES data is one that maximizes the amount of star formation occurring in small objects beyond current detection limits.

Next, in the middle panel of Figure \ref{fig:sfrd_recon}, we show the SFR-$M_h$ relationship. Because the SFE is ultimately degenerate with our model for halo mass growth rates, it is useful to simply look at the product $\SFR = \fstar \MAR$, which is insensitive to changes in the halo growth model. In this space, a flat SFE translates to $\SFR \propto M_h$, whereas the UVLF-only models show $\SFR \propto M_h^{5/3}$ (roughly) to arbitrarily low masses, consistent with simple feedback arguments.

Finally, in the right panel of Figure \ref{fig:sfrd_recon}, we show the reconstructed SFRD (blue contours), which shows a level of star formation at $z \sim 10$ that is $\sim 10$ times higher than UVLF-based inferences (green contours), even when those UVLFs are extrapolated to the atomic threshold. The integrated SFRD is $\sim 20-30$x higher than the SFRD inferred from bright objects ($\MUV \gtrsim -17$) only (dashed curves). However, we emphasize that such elevated values of the SFRD need not continue to redshifts $z \lesssim 15$ -- we simply have no mechanism in our model capable of stifling star formation differentially as a function of time.

\begin{figure}
\begin{center}
\includegraphics[width=0.49\textwidth]{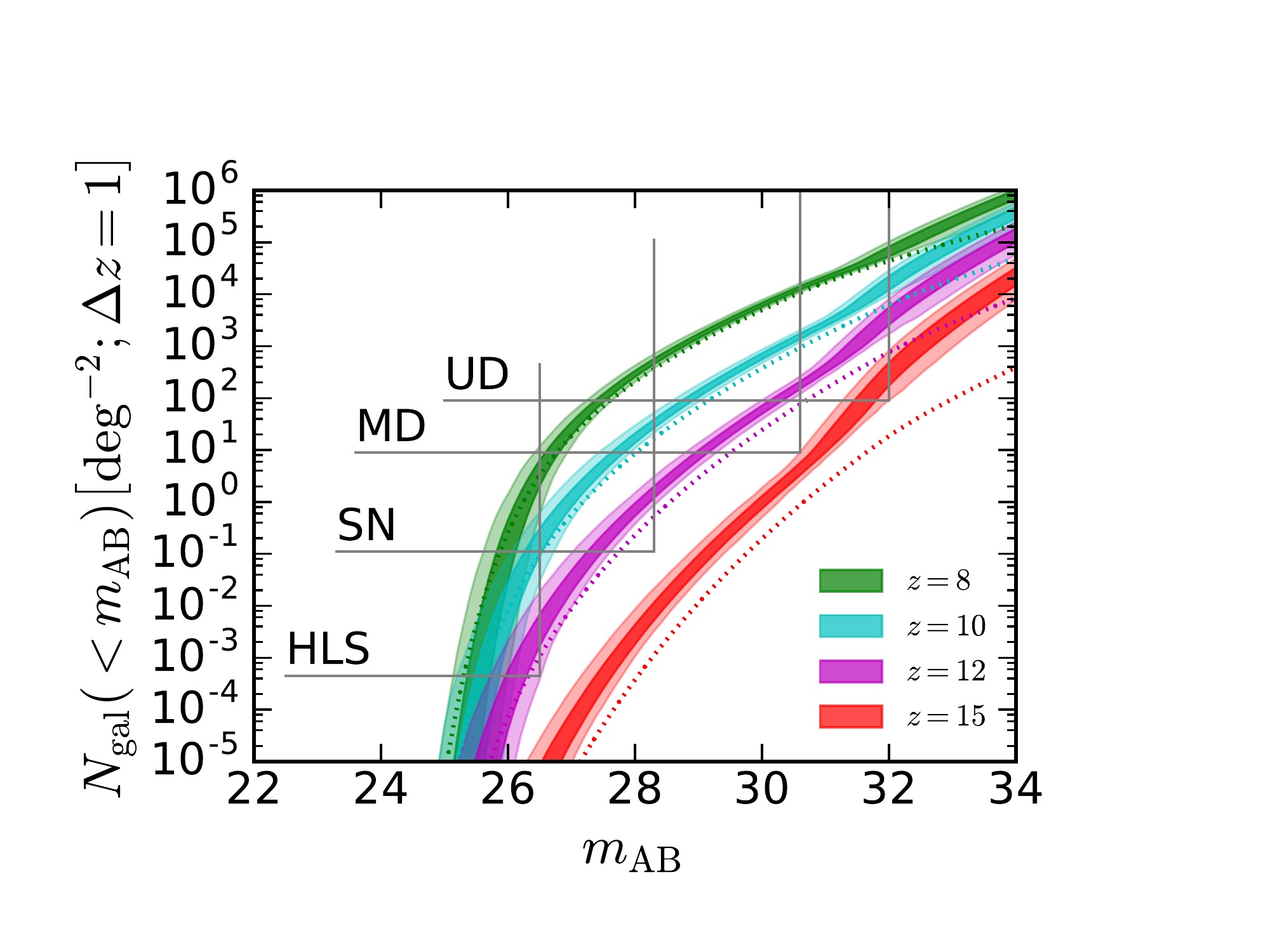}
\caption{{\bf Predictions for galaxy counts at high-$z$.} We show the cumulative surface density of galaxies \textit{brighter} than observed AB magnitude $m_{\mathrm{AB}}$ per unit redshift. Dotted curves show predictions based on the extrapolation of UVLFs, while the filled contours denote results obtained using UVLFs and the EDGES measurement. Four survey strategies are also shown (described in text), with vertical lines indicating sensitivity limits, and horizontal lines highlighting the limit at which a single galaxy is found in the search area. An ultra-deep survey with \jwst\ that detects $\sim 10$ or fewer galaxies at $z \sim 12$ would be strong evidence \textit{against} our model, or that even fainter sources are responsible for the EDGES signal.}
\label{fig:jwst}
\end{center}
\end{figure}

We have intentionally limited our model to the realm of atomic cooling halos, since they should be the most readily detectable objects in future deep surveys. In Figure \ref{fig:jwst}, we show our predictions in the context of several survey strategies, including ultra-deep (UD) and medium-deep (MD) surveys with JWST, and wide-field surveys with WFIRST, similar to their proposed supernova survey (SN) and high-latitude survey (HLS). We adopt the survey depths and areas quoted in \citet[][their \S3.3]{Mason2015}, with the exception of the SN survey, which is like their \jwst\ wide-field (WF) survey but covering 5x the area.

Our reference model, constructed without knowledge of the EDGES measurement, is shown in the dotted lines, and show a relatively pessimistic outlook for planned surveys at $z \gtrsim 10-15$. Such a model predicts that none of these surveys will find galaxies at $z \sim 15$, though 10-100 galaxies could be detected in each of the UD, MD, and SN surveys at $z \sim 12$.

The joint UVLF+EDGES results are shown with filled contours. If the faint-end of the UVLF truly steepens, an ultra-deep survey with JWST could see $\sim 100$ galaxies at $z \sim 12$, and perhaps $\sim 10$ at $z \sim 15$. Unfortunately, however, the scenario in which no $z\sim 15$ galaxies are detected is still consistent with our model at the $2 \sigma$ level. As a result, the strongest evidence of a steepening UVLF at high redshift would be the detection of more than $\sim 10$ galaxies at $z \sim 12$ in a JWST ultra-deep field. In other words, strong upper limits at $N_{\mathrm{gal}} < 10$ at $z \gtrsim 12$ would point toward even fainter sources as the driving force behind the EDGES signal. In this case, 21-cm observations \citep[with, e.g., HERA, SKA;][]{DeBoer2017,Koopmans2015} may be the only way to constrain the properties of these sources in any detail.

\section{Discussion} \label{sec:discussion}
\subsection{Implications}
As shown in the previous section, reconciliation of the EDGES measurement and high-$z$ UVLF constraints is not so easily attained. We find that evolution of the SFE both in its normalization and its shape are required, if restricted to changes in the properties of atomic-cooling halos and ``normal'' stellar populations.

The redshift evolution in the SFE we recover is not necessarily predicted from theoretical models. For example, while simple feedback arguments predict $\fstar \propto M_h^{2/3}$, as is roughly observed, they also predict that the normalization (at fixed mass) should \textit{increase} with $z$ as $\fstar \propto (1+z)^{1/2}$ or $\fstar \propto (1+z)$ \citep[e.g.,][]{Dayal2014,Furlanetto2017}. Physically, this rise comes from the increased binding energy of halos (at fixed mass) at higher redshifts, enhancing the resiliency of galaxies to supernova feedback. We recover precisely the opposite trend, roughly $\fstar \propto (1+z)^{-1}$. This could be an indication that our model over-estimates mass accretion rates at the highest redshifts, forcing the SFE to compensate. Indeed, our model \textit{must} overestimate inflow rates, as we have completely neglected mergers as a means by which galaxies grow. However, simulations suggest that mergers are subdominant at high-$z$ \citep[e.g.,][]{Goerdt2015}, i.e., unlikely to account for the factor of $(1+z)$ or more that is required here. Despite the decline in SFE with redshift, star formation rates (at fixed $M_h$) still rise with redshift (see Fig. \ref{fig:sfrd_recon}).

Perhaps the more surprising result is the need for a change in the shape of the SFE. In our fitting, the most extreme behavior that is allowed to occur is  $\fstar = \mathrm{constant}$ with respect to $M_h$. More gradual evolution in the slope of an unbroken power-law was a possible solution, as we allowed the slope of the SFE curve to evolve in time in our fits. Because the EDGES trough only persists to $z \sim 15$, its demand for elevated star formation ends below that point. In principle, the mechanisms that increase star formation in small galaxies could therefore also end at $z \sim 15$. We have chosen not to complicate our model by introducing an explicit parameterization like this. However, we note that even if this efficient star formation persists until $z \sim 6$, our models remain consistent with the \citet{Planck2015} constraints on the CMB optical depth. For an escape fraction $\fesc =0.1$, our models generate $\tau_e \sim 0.08$ including the faint-end UVLF steepening, and $\tau_e \sim 0.06$ in a model with an unbroken power-law in the SFE. As a result, $\fesc \lesssim 0.05$ is viable in models with enhanced star formation in halos $M_h \lesssim 10^{10} \ \Msun$.

This need for substantially enhanced star formation in low-mass objects could of course be ameliorated by minihalos, which are exceedingly abundant in the early Universe. We recently found that Population III stars and their remnants (the presumed inhabitants of minihalos) have only a subtle impact on the global signal in all but the most extreme cases \citep{Mirocha2018}. As a result, if number counts of galaxies at $z \sim 12-15$ remain low (see Figure \ref{fig:jwst}), it might imply rather extreme PopIII star formation. In this case, high-$z$ supernova surveys might also help distinguish the sources dominating high-$z$ star formation \citep{Mebane2018}.

It is also possible that the somewhat contrived evolution in the SFE we require to simultaneously fit UVLFs and the EDGES signal is signifying the emergence of new source populations, rather than revealing some unexpected evolution in the properties of star-forming galaxies. \citet{BoylanKolchin2017} recently pointed out that the UVLFs of globular clusters forming at high-$z$ could be comparable to those expected of galaxies themselves. This possibility is supported observationally as well, given that many objects in lensing fields are remarkably small \citep{Bouwens2017b}. Though the faint-end slope of a globular cluster population is shallower than we require \citep[at $\alpha\sim -1.7$;][]{BoylanKolchin2017}, lensing campaigns are biased toward their detection, and as a result, may obscure the true shape of the UVLF at high-$z$ \citep{Zick2018}. Moving forward, it will be vital to understand such biases well enough to distinguish intrinsically steep UVLFs from those that have been inflated through such effects.

New source populations may also be warranted if the radio background model is to remain a viable explanation of the signal's amplitude. We have found that if star-forming galaxies are to generate this background, they must produce low-frequency emissions $\sim 10^3$ more efficiently than galaxies today. To make matters worse, this epoch of enhanced radio emission must rapidly come to an end in order to both match the shape of the EDGES signal and to fall below the $z=0$ excess reported by ARCADE-2 \citep{Fixsen2011}. Accreting super-massive black holes may be a more viable source of this background \citep{EwallWice2018}, but, as in the case of radio emission from star formation, it remains unclear what mechanisms would strongly suppress  emissions at $z \lesssim 15$.

\subsection{Caveats}
Finally, we discuss the possible shortcomings of our approach, and if any assumptions or approximations we have made could account for the discrepancy between UVLF-calibrated predictions for the global 21-cm signal \citepalias{Mirocha2017} and the observed global 21-cm signal reported in \citet{Bowman2018}.

Our underlying model is similar to many others in the literature \citep[e.g.,][]{Sun2016,Mashian2016,Mason2015}, and agrees well with the form of the SFE inferred in these works as well as their predictions for UVLFs at higher redshifts. The core assumption in each model is that star-formation is fueled by the inflow of pristine gas, proceeding continuously such that halos of the same mass have identical assembly histories in each of their constituent components (i.e., total, gas, stellar, and metal masses). This model clearly cannot be correct in detail, as galaxies are known to exhibit scatter in their luminosities (at fixed mass), not to mention a slew of factors that can set galaxies of the same mass on entirely different long-term growth trajectories (i.e., not just minor excursions from an otherwise smooth history). However, generalizing our models to address such complexities will not obviously result in a systematic rise in star formation globally at $z > 10$, and thus may not reduce the need for efficient star formation in objects beyond current detection limits.

For example, scatter in the star formation rate (and thus luminosity) in halos of fixed mass is more likely to bias the SFE to high values, if in fact the scatter is lognormal. Furthermore, because galaxies spend some of their time with SFRs less than the expected rate (given $M_h$), it is difficult to dramatically change the star formation rate \textit{density}. As a result, our neglect of scatter results in a model with a conservative estimate of the SFRD.

One economical way to reduce the need for enhanced star formation in objects beyond current detection limits is to invoke obscuration in the objects we do see. Though we perform a standard dust correction to correct for reddening, we have neglected the possibility that some fraction of emission in UV-detected galaxies could be completely extincted \citep[e.g.,][]{Bowler2018}. The right panel of Fig. 4 shows that the required excess SFRD at $z \sim 18$ is more than an order of magnitude beyond what is possible with galaxies of similar mass to those observed at $z \sim 8-10$, which would in any case require very extreme dust corrections to completely resolve. Our models also broadly agree with constraints on the stellar mass functions at high-$z$ \citep{Stefanon2017}, which are less susceptible to dust effects, so any bias in our inferred SFE is likely small.

The semi-empirical nature of our model permits models that could reasonably be considered contrived. However, the idea that feedback ceases to regulate star formation in sufficiently small halos -- one interpretation of a flat SFE -- is not entirely unappealing. For example, the dynamical timescale approaches the lifetimes of massive stars in halos $M_h \lesssim 10^9 \ \Msun$ \citep{FaucherGiguere2018}. This, coupled with rapid inflow rates, could allow galaxies to form stars out of new gas before the previous generation of stars has a chance to drive proto-stellar material away via supernovae blast-waves. Alternatively, persistent contrivances may point to a more fundamental failure of models based on abundance matching and smooth inflow-driven star formation. 21-cm observations could thus provide constraints on philosophically different approaches to galaxy evolution modeling \citep[e.g.,][]{Kelson2016}.

Finally, our results demonstrate the importance of multi-wavelength measurements of the high-$z$ universe for rigorously constraining the properties of luminous sources during the cosmic dawn. Ironically, early predictions for the 21-cm global signal \citep[e.g.,][]{Furlanetto2006,Pritchard2010a,Mesinger2013,Mirabel2011,Mirocha2015,Fialkov2014b} yielded absorption troughs at $\lesssim 80$~MHz in their fiducial treatments. These models, however, were not tied to observations of the galaxy population, simply assuming a single mean star formation efficiency across all halos so could not be taken as any more than qualitative predictions. The dramatic improvements in measurements of the galaxy LF at $z>6$ over the past several years \citep[e.g.,][]{Vanderburg2010,McLure2013,Finkelstein2015,Bouwens2015,McLeod2016,Livermore2017,Oesch2017,Bouwens2017} required a recalibration of our expectations: like nearby galaxy populations, they require a declining star formation efficiency in small halos -- precisely those that are abundant at even higher redshifts -- and this understanding allowed us to make \emph{quantitative} predictions for their effects on the 21-cm background. 

Thus it is only by \emph{combining} UVLF measurements with the EDGES signal that we can understand the puzzle the latter poses for galaxy formation physics. Models that accurately describe the physics driving both measurements are essential for extracting the most useful and compelling implications, and improvements to the underlying theoretical models and their predictive range (including allowances for sources we have ignored in the present treatment, such as Population~III star formation) will be necessary as galaxy surveys, the 21-cm data, and other measurements of the cosmic dawn improve over the next several years.

\section{Conclusions} \label{sec:conclusions}
We summarize our main findings as follows:
\begin{itemize}
    \item A 78 MHz feature in the global 21-cm signal is not expected based on the extrapolation of UVLFs, as outlined in \citetalias{Mirocha2017}. In order to induce the EDGES feature without violating UVLFs at $z \lesssim 10$, we must appeal to enhanced star formation in objects currently beyond detection limits. This will remain the case even in the event that the amplitude of the signal is revised downward in future studies, provided that its central frequency remains intact. 
    \item We find that, if the efficiency of star formation tends to a constant $\sim$ few percent in halos $M_h \lesssim 10^{10} \Msun$, galaxies in atomic cooling halos alone can provide enough star formation to explain the timing of the EDGES measurement. This implies a corresponding steepening in the UVLF at high-$z$ relative to past predictions, which could be tested by forthcoming observations. Detection of fewer than $\sim 10$ galaxies at $z \sim 12$ in a JWST ultra-deep field would be compelling evidence that even fainter sources are required to explain the EDGES measurement, in which case 21-cm observations with interferometers like HERA, LOFAR, and the SKA, may be the only way to constrain the properties of the first luminous sources in any detail.
    \item We find that $f_X \gtrsim 10$ is also required to fit the EDGES signal. Given that we have assumed an unabsorbed X-ray spectrum, the true value could be even higher.
    \item In order for the first sources to generate a radio background capable of amplifying the global 21-cm signal, they must produce radiation at $\sim 1-2$ GHz some $\sim 10^3$ times more efficiently (per unit SFR) than star-forming galaxies today. Furthermore, this efficient radio emission must terminate beyond $z \sim 15$ in order to both match the shape of the signal measured by EDGES and to fall below the $z=0$ excess.
    \item The excess cooling model only removes the need for substantial radio emission from high-$z$ sources -- the implications for the UVLF remain unchanged, as the timing of the EDGES signal, and its sharpness, provide a vital constraint on early galaxies. 
\end{itemize}

The authors thank Rick Mebane, Raul Monsalve, Charlotte Mason, Aaron Ewall-Wice, and Louis Abramson for many useful discussions. This work was supported by the National Science Foundation through award AST-1636646 and by NASA through award NNX15AK80G. In addition, this work was directly supported by the NASA Solar System Exploration Research Virtual Institute cooperative agreement number 80ARC017M0006. We acknowledge support from a NASA contract supporting the ``WFIRST Extragalactic Potential Observations (EXPO) Science Investigation Team'' (15-WFIRST15-0004), administered by GSFC. This work used computational and storage services associated with the \textsc{Hoffman2} Shared Cluster provided by UCLA Institute for Digital Research and Education's Research Technology Group, and relied on the Python packages \textsc{numpy}\footnote{\url{http://www.numpy.org/}} \citep{Oliphant2007} and \textsc{matplotlib}\footnote{\url{http://www.matplotlib.org/}} \citep{Hunter2007}.

\bibliography{references}
\bibliographystyle{mn2e_short}

\end{document}